\documentclass[conference]{IEEEtran}

\usepackage{cite}
\usepackage{amsmath,amssymb,amsfonts}
\usepackage{algorithmic}
\usepackage{graphicx}
\usepackage{textcomp}
\usepackage{xcolor}

\usepackage[caption=false]{subfig}
\usepackage{harveyballs}
\usepackage{tabularx,colortbl}
\usepackage{array}
\usepackage{multirow}
\usetikzlibrary{calc, arrows.meta,arrows}
\usetikzlibrary{positioning}
\usepackage[inline]{enumitem}
\usepackage{flushend}

\newcommand{\name}[1]{BiTHaC}

\usepackage{xcolor}

\begin{document}

\makeatletter
\def\ps@IEEEtitlepagestyle{%
\def\@oddfoot{\parbox{\textwidth}{\footnotesize
Author's version of a paper accepted for publication in Proceedings of the 2025 IEEE 50th Conference on Local Computer Networks (LCN). 
\\
\textcopyright{} 2025 IEEE. 
Personal use of this material is permitted.  
Permission from IEEE must be obtained for all other uses, in any current or future media, including reprinting/republishing this material for advertising or promotional purposes, creating new collective works, for resale or redistribution to servers or lists, or reuse of any copyrighted component of this work in other works.\vspace{1.2em}}
}%
}
\makeatother

\title{Bidirectional TLS Handshake Caching for Constrained Industrial IoT Scenarios}

\author{\IEEEauthorblockN{%
Jörn Bodenhausen\IEEEauthorrefmark{1}, 
Simon Mangel\IEEEauthorrefmark{1}, 
Thomas Vogt\IEEEauthorrefmark{1},
Martin Henze\IEEEauthorrefmark{1}\IEEEauthorrefmark{4}
}
\IEEEauthorblockA{%
\IEEEauthorrefmark{1}\textit{Security and Privacy in Industrial Cooperation}, RWTH Aachen University, Germany \\
\IEEEauthorrefmark{4}\textit{Cyber Analysis \& Defense}, Fraunhofer FKIE, Germany\\
\{bodenhausen, vogt, henze\}@spice.rwth-aachen.de \(\cdot\) simon.mangel@rwth-aachen.de 
}
}

\maketitle

\begin{abstract}
While TLS has become the de-facto standard for end-to-end security, its use to secure critical communication in evolving industrial IoT scenarios is severely limited by prevalent resource constraints of  devices and networks.
Most notably, the TLS handshake to establish secure connections incurs significant bandwidth and processing overhead that often cannot be handled in constrained environments.
To alleviate this situation, we present \name{} which realizes bidirectional TLS handshake caching by exploiting that significant parts of repeated TLS handshakes, especially certificates, are static.
Thus, redundant information neither needs to be transmitted nor corresponding computations performed, saving valuable bandwidth and processing resources.
By implementing \name{} for wolfSSL, we show that we can reduce the bandwidth consumption of TLS handshakes by up to 61.1\% and the computational overhead by up to 8.5\%, while incurring only well-manageable memory overhead and preserving the strict security guarantees of TLS.
\end{abstract}


\section{Introduction} \label{sec:intro}

With the omnipresence of Internet technology in modern communication, its use in industrial settings such as smart cities~\cite{kim2017smart}, vehicular communication~\cite{zhang2020vehicle}, sensor networks \cite{ko2011industry}, or industrial control systems~\cite{lenz2025cofacs} becomes increasingly relevant. 
Due to the connection to critical infrastructure and the resulting significance of operational safety~\cite{rondon2020poisonivy}, end-to-end (E2E) security of industrial communication is crucially important~\cite{wagner2024madtls}.

To ensure interoperability, Transport Layer Security (TLS), the de-facto standard protocol for E2E security on the Internet, is a key candidate for securing IIoT communication~\cite{dahlmanns2022tls}.
Often, the use of TLS in industrial communication is even required by laws and standards~\cite{rademacher2022bounds,heimgaertner2018distributed}.
However, as any security solution in constrained IIoT scenarios, TLS has to adhere to often strict constraints regarding bandwidth, processing, and memory prevalent across industrial devices and networks~\cite{rfcConstrained,henze2016cppl}, especially when considering the push towards deeply embedded devices relying on wireless communication~\cite{bodenhausen2023securing, rademacher2021path}. 
Contrary to these strict resource bounds, especially the use of public-key cryptography during TLS handshakes incurs significant bandwidth and processing overheads. 
This makes the use of TLS in constrained IIoT scenarios challenging~\cite{rademacher2022bounds,neto2016aot}, an issue that will further exaggerate with the ongoing transition towards post-quantum cryptography~\cite{Schwabe2021More,bang2022iot}.

To account for the restrictions imposed by constrained IIoT scenarios and still allow for the use of TLS, related work has proposed optimizations for TLS:
\begin{enumerate*}[label=(\roman*)]
\item use of profiles~\cite{Gupta2005Sizzle,Tschofenig2016Transport} and optimized encoding~\cite{Raza20126LoWPAN,Rescorla2023Compact},
\item out-of-band transmission~\cite{Kothmayr2012DTLS,Wouters2014Using} and handshake delegation~\cite{Hummen2014Delegation,Raza2016S3K}, as well as
\item caching of sessions~\cite{Eronen2008Transport,Tange2020rTLS} and handshakes~\cite{Apostolopoulos1999Transport,Santesson2016Transport}. 
\end{enumerate*}

Especially caching allows for substantial resource savings. %
However, \emph{session} caching~\cite{Eronen2008Transport,Tange2020rTLS} requires frequent full handshakes to cryptographically decouple connections~\cite{hebrok2023we}, making it unsuited for industrial scenarios~\cite{rademacher2022bounds}.
Contrary, \emph{handshake} caching~\cite{Apostolopoulos1999Transport,Santesson2016Transport} fully preserves E2E security guarantees and thus is an extremely promising candidate to optimize TLS for constrained IIoT scenarios.
Still, existing approaches mainly target to reduce the bandwidth overhead imposed by server certificates in Internet communication and thus do not consider the specifics of the constrained IIoT.

To fill this gap, we present \name{}, our approach to realize bidirectional TLS handshake caching for constrained IIoT scenarios.
\name{} exploits that significant parts of the TLS handshake are static, and thus cacheable.
Consequently, after an initial full handshake, those parts neither need to be transmitted (saving bandwidth) nor corresponding computations performed (reducing processing overhead and latency) again.
To this end, \name{} adapts the idea of caching server certificates on the client~\cite{Santesson2016Transport} and adds a corresponding counterpart to cache client certificates on the server. %
Furthermore, \name{} specifically accounts for constrained devices through memory-optimized caching and skipping of redundant computations.
As such, \name{} substantially decreases the overhead of TLS handshakes without impacting E2E security guarantees.

\noindent\textbf{Contributions.}
We address the need to reduce bandwidth consumption and computational overhead of the TLS handshake in constrained scenarios with the following contributions:
\begin{enumerate}
    \item We tailor the idea of caching static parts of server messages in TLS handshakes to constrained IIoT scenarios and provide a mechanism to also cache static parts of client messages without compromising security.
    \item We propose a novel paradigm to handling client-side caches such that redundant yet resource-intensive computations can be skipped with minimal memory impact.
    \item Our evaluation for TLS 1.2 and 1.3 shows that \name{} successfully reduces the bandwidth consumption by up to 61.1\%. %
    Likewise, \name{} reduces the computational overhead of TLS handshakes by up to 8.5\%.
\end{enumerate}

\noindent\textbf{Availability Statement.} Our implementation of \name{} is available at https://github.com/RWTH-SPICe/BiTHaC

\section{E2E Security in Constrained Environments} \label{sec:background}

To better understand the root causes for the challenges of using E2E security and specifically TLS in resource-constrained environments \cite{bodenhausen2023securing,rademacher2022bounds,wagner2024madtls}, we first introduce TLS and then discuss resulting implications for constrained environments.

\subsection{Transport Layer Security (TLS)} \label{sec:back:tls}

TLS is an application layer cryptographic protocol to realize communication security in an E2E manner, i.e., between client and server.
Over the years, multiple updates and extensions of TLS have been developed, with TLS 1.2 \cite{rfcTLS12} and 1.3 \cite{rfcTLS13} being the currently most relevant ones~\cite{dahlmanns2022tls}. 

TLS realizes its functionality through various sub-protocols. 
Both for security and resource-consumption, the \emph{handshake protocol} for establishing a secure connection is particularly relevant. 
It is used to negotiate the required parameters and authenticate one or both peers. %
Negotiation allows the protocol to be versatile and support a wide range of cryptographic algorithms. 
Moreover, an extension mechanism can be used to realize additional functionality. 
Using this mechanism, the client can offer additional functionality which the server can choose to accept. 
The extension mechanism is designed rather lightweight and consist of a header with a two-byte identifier, a two-byte length field, followed by extension specific data.

\begin{figure}[t]
    \begin{center}
        
      \resizebox{\columnwidth}{!}{\begin{tikzpicture}
            \node[above] at (0,0) {Message Sizes};

            \draw[] (-1.5,0) -- (1.5,0);
            \draw[] (0,0) -- (0,-4.7);
            \draw[] (-1.5,0) -- (-1.5,-4.7);
            \draw[] (1.5,0) -- (1.5,-4.7);
            \draw[] (-1.5,-4.4) -- (1.5,-4.4);
            \draw[] (-1.5,-4.7) -- (1.5,-4.7);

            \def\vOL{-6};
            \def\vIL{-2}; 
            \def\vML{-4};

            \node[above] at (\vML,0.5) {\textbf{TLS 1.2 Handshake}};

            \draw (\vOL,0) -- (\vOL,-4.6);
            \node[above right] at (\vOL,0) {\underline{Client}};

            \draw (\vIL,0) -- (\vIL,-4.6);
            \node[above left] at (\vIL,0) {\underline{Server}};

            \node[right] at (\vOL, -0.3) {Client Hello};
            \node[left] at (0, -0.3) {130 B};

            \draw[-{Stealth[scale=1.5]}] (\vOL,-0.5) -- (\vIL,-0.5);

            \node[left] at (\vIL, -0.9) {Server Hello};
            \node[left] at (0, -0.9) {91 B};

            \node[left] at (\vIL, -1.2) {Certificate};
            \node[left] at (0, -1.2) {\textbf{796 B}};

            \node[left] at (\vIL, -1.5) {Server Key Exchange};
            \node[left] at (0, -1.5) {338 B};

            \node[left] at (\vIL, -1.8) {Certificate Request};
            \node[left] at (0, -1.8) {35 B};

            \node[left] at (\vIL, -2.1) {Server Hello Done};
            \node[left] at (0, -2.1) {9 B};

            \draw[{Stealth[scale=1.5]}-] (\vOL,-2.3) -- (\vIL,-2.3);

            \node[right] at (\vOL, -2.7) {Certificate};
            \node[left] at (0, -2.7) {\textbf{790 B}};

            \def\vL32{}
            \node[right] at (\vOL, -3) {Client Key Exchange};
            \node[left] at (0, -3) {75 B};

            \node[right] at (\vOL, -3.3) {Certificate Verify};
            \node[left] at (0, -3.3) {269 B};

            \def\vL34{}
            \node[right] at (\vOL, -3.6) {Finished};
            \node[left] at (0, -3.6) {45 B};

            \draw[-{Stealth[scale=1.5]}] (\vOL,-3.8) -- (\vIL,-3.8);

            \node[left] at (\vIL, -4.2) {Finished};
            \node[left] at (0, -4.2) {45 B};

            \draw[{Stealth[scale=1.5]}-] (\vOL,-4.4) -- (\vIL,-4.4);

            \def\vOR{6};
            \def\vIR{2}; 
            \def\vMR{4};

            \draw (\vIR,0) -- (\vIR,-4.6);
            \node[above right] at (\vIR,0) {\underline{Client}};

            \draw (\vOR,0) -- (\vOR,-4.6);
            \node[above left] at (\vOR,0) {\underline{Server}};

            \node[above] at (\vMR,0.5) {\textbf{TLS 1.3 Handshake}};
            \def\vOffs{1.5};

            \node[right] at (\vIR, -0.3) {Client Hello};
            \node[left] at (\vOffs, -0.3) {229 B};

            \draw[-{Stealth[scale=1.5]}] (\vIR,-0.5) -- (\vOR,-0.5);

            \node[left] at (\vOR, -0.9) {Server Hello};
            \node[left] at (\vOffs, -0.9) {128 B};

            \node[left] at (\vOR, -1.2) {Encrypted Extensions};
            \node[left] at (\vOffs, -1.2) {28 B};

            \node[left] at (\vOR, -1.5) {Certificate Request};
            \node[left] at (\vOffs, -1.5) {75 B};

            \node[left] at (\vOR, -1.8) {Certificate};
            \node[left] at (\vOffs, -1.8) {\textbf{816 B}};

            \node[left] at (\vOR, -2.1) {Certificate Verify};
            \node[left] at (\vOffs, -2.1) {286 B};

            \node[left] at (\vOR, -2.4) {Finished};
            \node[left] at (\vOffs, -2.4) {74 B};

            \draw[{Stealth[scale=1.5]}-] (\vIR,-2.6) -- (\vOR,-2.6);

            \node[right] at (\vIR, -3.1) {Certificate};
            \node[left] at (\vOffs, -3.1) {\textbf{810 B}};

            \node[right] at (\vIR, -3.4) {Certificate Verify};
            \node[left] at (\vOffs, -3.4) {286 B};

            \node[right] at (\vIR, -3.7) {Finished};
            \node[left] at (\vOffs, -3.7) {74 B};

            \draw[-{Stealth[scale=1.5]}] (\vIR,-4) -- (\vOR,-4);
            
            \node[left] at (0, -4.55) {\textbf{2.578 B}};
            \node[left] at (\vOffs, -4.55) {\textbf{2.806 B}};

        \end{tikzpicture}}
    \end{center}
    \vspace{-0.8em}
    \caption{Handshake comparison between TLS 1.2 and 1.3 alongside measured message sizes. Both handshakes use mutual authentication with single self-signed 2048-bit RSA certificates. A single cipher is offered by the client, resulting in an ECDHE key exchange and use of AES 256 GCM SHA384.}
    \label{fig:HS} 
\end{figure}
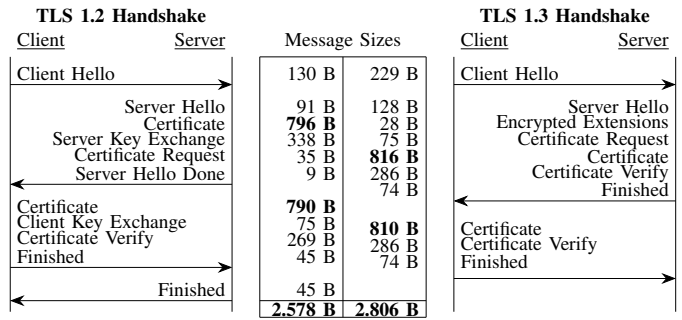

While previous version updates kept the handshake largely unchanged, TLS 1.3 introduces notable changes to achieve a more secure and faster connection establishment. 
The main differences are illustrated in Fig.\ \ref{fig:HS} alongside the size of each message based on an exemplary connection establishment. 
In TLS 1.2, connection establishment requires four flights and starts with the \textit{Hello} messages, in which fundamental parameters and the extensions are negotiated. 
Subsequently, key-exchange and authentication steps are performed through the exchange of \textit{Certificate} and \textit{Key Exchange} messages. 
Finally, the connection establishment is concluded through \textit{Finished} messages. 
Connection establishment in TLS 1.3 differs substantially and requires only three flights. 
To achieve this, dedicated \textit{Key Exchange} messages are omitted and their functionality moved to extensions. 
Moreover, all messages after the \textit{Hello} messages are encrypted, including a newly added \textit{Encrypted Extensions} message, for additional privacy. 

The message sizes in Fig.\ \ref{fig:HS} show that a single handshake adds up to a bandwidth overhead in the order of KBs, even for single self-signed certificates.
In reality, chains of multiple certificates further increase the bandwidth overhead \cite{kampanakis2022faster}.
Nevertheless, even in this simple case, the two \textit{Certificate} messages make up 61.5\% of the TLS 1.2 handshake messages and 57.9\% for TLS 1.3. 
Moreover, when considering the upcoming use of post-quantum algorithms, this will further increase, as the utilized public keys are several KBs in size \cite{kampanakis2022faster} and expected to be utilized in a hybrid manner \cite{ietf-tls-hybrid-design-10}.

\subsection{Constrained Industrial IoT Scenarios} \label{sec:background:constrained}

Devices and networks in Industrial IoT (IIoT) environments impose various resource constraints~\cite{rfcConstrained,bodenhausen2023securing,eggert2014sensorcloud}.
Most notably, industrial devices and networks are severely constrained w.r.t.\ bandwidth, processing,  memory, or energy~\cite{rfcConstrained,serror2021challenges,karlof2004tinysec,henze2013maintaining}, especially for wireless communication~\cite{bodenhausen2023securing,michaelides2025industry5G,brighente2022interference}.

IIoT \emph{devices} are often limited w.r.t.\ processing power, memory, and energy \cite{rfcConstrained}. 
Using TLS to secure communication affects all three dimensions. 
Energy consumption is affected, as the transmission of larger messages over wireless networks consumes considerable power. 
Moreover, asymmetric cryptography utilized in TLS is particularly resource intensive \cite{restuccia2020low}, leading to substantial computational and indirectly power consumption overhead. 
Lastly, the memory overhead of TLS, especially the impact of asymmetric cryptography \cite{restuccia2020low}, is challenging for devices with only a few KB of memory \cite{rfcConstrained}. 

Considering IIoT \emph{networks}, constraints resulting from the prevalent use of wireless network technology limit the bandwidth available to each device and thus its ability to establish and maintain a TLS connection \cite{rademacher2022bounds}. 
Moreover, wireless network technologies commonly impose restrictions on individual message sizes \cite{bodenhausen2024adaptive}. 
For instance, LoRaWAN has a maximum message size of only 256 bytes \cite{bodenhausen2023securing}, which would already cause the \textit{Certificate} messages from our example with a single certificate (cf.\ Fig.\ \ref{fig:HS}) to be fragmented, thus invoking additional overhead on lower network layers and delays. 

Overall, the use of certificate chains and associated computations for mutual authentication are integral for security in TLS.
At the same time, they heavily influence the overall resource overhead and particularly bandwidth consumption, especially in constrained IIoT environments, often rendering the application of secure communication infeasible.

\section{Related Work on Optimizing TLS Overhead} \label{sec:rel_work}

Various work motivate the use of TLS in the IoT~\cite{kothmayr2011poster,brachmann2012end,behrens2017internet} without detailing applicability and resulting overheads. %
Other proposals suggest to realize E2E secure communication other than TLS, e.g., based on HIP~\cite{sahraoui2014compressed,sahraoui2015efficient} or custom solutions~\cite{granjal2017adaptable,plusquellic2023privacy}.
As these jeopardize the interoperability of E2E security and often do not meet the requirements of industrial contexts~\cite{rademacher2022bounds,heimgaertner2018distributed}, our discussion focuses on approaches to optimize the overhead imposed by TLS.

\textbf{Profiles \& Optimized Encoding} aim to reduce overhead through enhanced message formats or protocol restrictions, leaving protocol features and security guarantees largely unchanged. 
For example, header compression improves the overhead of Datagram TLS (DTLS) \cite{Raza20126LoWPAN,Raza2013Lithe,Banerjee2017eeDTLS}.
Moreover, Compact TLS 1.3 reduces the amount of transmitted information through re-design and a templating mechanism \cite{Rescorla2023Compact}. 
However, such savings usually come at the cost of interoperability. 
Improvements on the typically used X.509 \cite{Boeyen2008Internet} certificates encoding are explored through compression \cite{McGrew2010Compressed,Ghedini2020TLS} and more efficient encoding \cite{OrtizYepes2015Optimizing,Hoeglund2020PKI4IoT,Mattsson2023CBOR}.
Still, these improved certificates still need to be transmitted for every connection. 
Conversely, different approaches reduce the overhead of TLS through restriction to a subset of available features based on profiles, e.g., exclusively using ECC \cite{OrtizYepes2015Optimizing,Gupta2005Sizzle,Jung2009SSL}. 
Furthermore, an official set of profiles for IoT environments \cite{Tschofenig2016Transport} is being updated for TLS 1.3 \cite{Tschofenig2023TLS/DTLS}. 
As such configuration optimizations are relevant in any particular scenario, they are orthogonal to more fundamental improvements.

\textbf{Out-of-Band Transmission} improves the overhead of TLS by utilizing an additional, typically unconstrained, communication channel. 
A prominent example is on-demand retrieval, as utilized by client certificate URLs \cite{3rd2011Transport}.
However, this mechanism merely shifts overhead from client to server and comes with various security considerations. 
As an alternative, pre-provisioning of such information allows to entirely leave out information in the actual connection establishment. 
Examples are the templating mechanism of Compact TLS 1.3 \cite{Rescorla2023Compact}, the direct use of raw public keys  \cite{Wouters2014Using}, or pre-sharing of symmetric keys \cite{Kothmayr2012DTLS}. 
However, as such information needs to be exchanged in advance, potential application scenarios and scalability are limited. 
Lastly, handshake delegation can be utilized for devices that are too constrained to establish a connection on their own by (partly) offloading connection establishment to another device \cite{Fouladgar2006Tiny,Polk2007Server,Hummen2014Delegation,Raza2016S3K}.
Such mechanisms, however, require the presence of a trustworthy powerful device~\cite{henze2014trustpoint}, which often is an unrealistic assumption. 
Finally, the less intrusive approach of certificate pre-validation on gateways \cite{Hummen2013Towards} avoids the unnecessary validation of invalid certificates, but still causes the full overhead for valid certificates (which is the majority of cases). 

\textbf{Caching} \label{sec:rel_work-caching}
omits static parts of the connection establishment across repeated connections. 
As the most drastic form of caching, session caching or resumption caches the entire established connection. %
Up to TLS 1.2, the used ID-based mechanism  \cite{rfcTLS12,Gupta2005Sizzle,Sobh2008Performance}, shows significant security implications, especially a lack of perfect forward secrecy (PFS). 
In contrast, session ticket-based resumption \cite{Eronen2008Transport} allows for PFS \cite{rfcTLS13} and proven advantageous for IoT environments \cite{Hummen2013Towards,Tschofenig2016Transport}.
Further extensions have been proposed, e.g., handshake delegation \cite{Hummen2013Extended,Hummen2014Delegation} and rTLS \cite{Tange2020rTLS}.
However, session resumption generally cryptographically links subsequent connections to the original connection, which limits their lifetime and leads to the need for frequent full handshakes~\cite{hebrok2023we}. 

\begin{figure*}[ht!]
\centering
\begin{minipage}{0.68\textwidth}
\vspace{6pt}
\subfloat[original TLS handshake\label{fig:design-overview:original}]{%
  \includegraphics[width=0.5\linewidth]{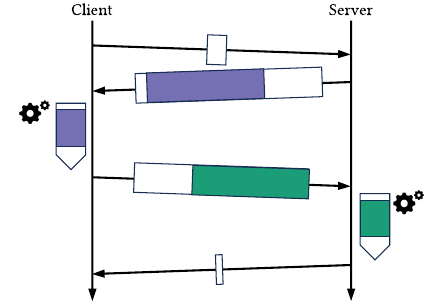}%
}\hfill
\subfloat[adapted \name{} handshake\label{fig:design-overview:ours}]{%
  \includegraphics[width=0.5\linewidth]{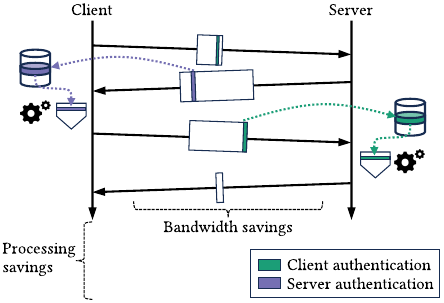}%
}
\caption{Significant parts of bandwidth- and computationally-intensive TLS handshakes are static (Fig.~\ref{fig:design-overview:original}). \name{} caches these parts to significantly reduce bandwidth consumption and processing overhead (Fig.~\ref{fig:design-overview:ours}).}
\label{fig:design-overview}
\end{minipage}%
\hfill
\begin{minipage}{0.3\linewidth}
  \includegraphics[width=\linewidth]{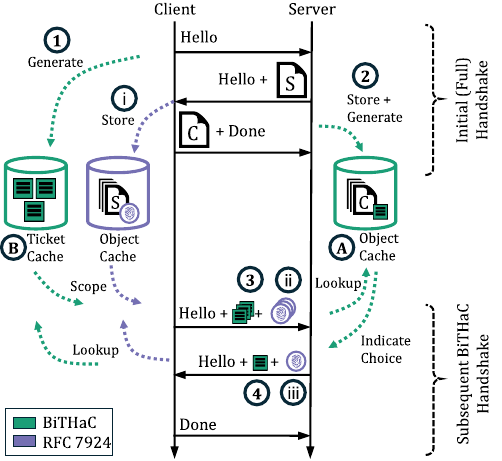}
  \caption{
  	  \name{} introduces a ticket-based signaling flow to realize server-side caching.
  }
  \label{fig:signaling}
\end{minipage}
\end{figure*}

Handshake caching constitutes a promising alternative (or complement) to session resumption by more granularly retaining aspects of the connection establishment rather than the entire connection. 
Various approaches for handshake caching were proposed \cite{Apostolopoulos1999Transport,Shacham2004Client,Langley2010Transport,OrtizYepes2015Optimizing}, leading to the standardization of the cached information extension (RFC 7924) \cite{Santesson2016Transport}.
Unlike other alternatives, this extension does not break the protocol and can be flexibly negotiated, positioning it as a promising approach to improve resource efficiency of the TLS handshake.
However, RFC 7924 \cite{Santesson2016Transport} still has significant drawbacks for constrained IIoT environments:
As caching only happens on the client-side, only half of the optimization potential is leveraged in IIoT scenarios where mutual authentication is imperative~\cite{schukat2014securing}.
Furthermore, device constraints are not considered, resulting in optimization potential w.r.t. the implementation of the cache. 
On the technical side, RFC 7924 does not work with TLS 1.3~\cite{Schwabe2021More} due to ephemeral content in \textit{Certificate} and \textit{Certificate Request} messages. %

\section{Bidirectional TLS Handshake Caching} \label{sec:design}

Existing approaches to optimize E2E security do not consider the specifics of IIoT scenarios (cf.\ Sec.\ \ref{sec:rel_work}), most importantly strict resource constraints on the client and the mandatory use of mutual authentication~\cite{schukat2014securing}.
To fill this gap, we present \name{} to optimize TLS for IIoT scenarios without compromising security.
As illustrated in Fig.\ \ref{fig:design-overview}, \name{} leverages the substantial static parts of TLS handshakes (Fig.~\ref{fig:design-overview:original}) to cache their transmission and corresponding processing (Fig.~\ref{fig:design-overview:ours}) without impacting E2E security.

To reduce the bandwidth overhead of TLS, we first leverage an existing caching mechanism to avoid the unnecessary transmission of static information sent by the server (Sec.\ \ref{sec:des:client}), before we introduce a novel caching scheme for static information sent by the client (Sec.\ \ref{sec:des:server}).
Based on this, we propose a mechanism to substantially reduce processing overhead with cached information (Sec.\ \ref{sec:des:cache}).
Finally, we present required adaptations for TLS 1.3 (Sec.\ \ref{sec:des:tls13}).

\subsection{Caching of Static Server Information (RFC 7924)} \label{sec:des:client}

To cache static server information on the client-side, we leverage the caching mechanism of RFC 7924 \cite{Santesson2016Transport} (illustrated in Fig.\ \ref{fig:signaling} in purple) that allows for caching of the server's \textit{Certificate} and \textit{Certificate Request} messages. 
After an initial full handshake, the client caches these messages (Fig.~\ref{fig:signaling}-i). 
In subsequent handshakes, it provides the server with options from its cache by including a list of object type and fingerprint (i.e., hash of the cached message) in the \textit{Client Hello} (Fig.~\ref{fig:signaling}-ii).
The server can then reuse cached elements by indicating its selection in the \textit{Server Hello} (Fig.~\ref{fig:signaling}-iii). 
Subsequently, the respective messages are replaced with the fingerprint and thus substantial bandwidth savings realized. 

As RFC 7924~\cite{Santesson2016Transport} only defines the changes to the TLS handshake protocol, to actually use it in \name{}, we derive a cache structure with two distinct lookup mechanism specifically tailored to constrained IIoT devices.  
Here, the primary lookup utilizes the type and fingerprint value to retrieve a cached object and can be used to access the required cached information selected by the server. 
Moreover, we utilize a secondary lookup via a peer index to scope the selection of cached elements included in the \textit{Client Hello} to each server.
Otherwise, the client would need to send \emph{all} cached elements for each new connection, leading to unnecessary bandwidth overhead and a potential risk of tracking. 
To uniquely identify servers, we rely on the commonly used server name indication (SNI) extension or the IP address, if SNI is not used.

With these adaptations, we realize the first half of bandwidth savings while catering for the specific requirements of the IIoT.

\subsection{A Novel Caching Scheme for Static Client Information} \label{sec:des:server}

While caching of static server information already saves some bandwidth (by caching the servers' \textit{Certificate} and \textit{Certificate Request} messages), the prevalent use of mutual authentication in constrained IIoT scenarios~\cite{schukat2014securing} coupled with tight bandwidth restrictions demand for also caching static client information on the server-side.
To this end, we propose a ticket-based mechanism for caching static client information, i.e., the \textit{Certificate} message (while extensible to other objects, we only consider the \textit{Certificate} message here).
Our design specifically focuses on bandwidth efficiency and client privacy, i.e., tracking prevention.
In the following, we provide an overview of our caching scheme, detail cache structure, extension format, ticket generation, and adapted behavior during the TLS handshake alongside Fig.\ \ref{fig:signaling} (in green).

\textbf{Overview of Caching Scheme.}
Our caching scheme for static client information relies on tickets to negotiate the use of cached information.
More specifically, the client generates a ticket which is used to identify an object cached at the server after a successful handshake (Fig.~\ref{fig:signaling}-1).
Analogously, the server stores the respective object and generates the associated ticket (Fig.~\ref{fig:signaling}-2).
In subsequent handshakes, the client includes suitable tickets in its \textit{Hello} message (Fig.~\ref{fig:signaling}-3).
If the server has a matching entry in its cache, it informs the client which cached information it selected for use (Fig.~\ref{fig:signaling}-4).
Subsequently, the client omits the selected cached information from sent messages.
Unlike for caching static server information, we thus move the choice which cached objects will be used to the entity storing the cached information (i.e., the server). 
This is necessary, as servers usually cannot identify clients at the beginning of connection establishment (while clients can easily identify servers, e.g., based on their SNI).
Consequently, if the client were to make the selection of which cached information to use (similar to RFC 7924), the server would need to send information on its \emph{complete} cache to the client, which is not feasible and would raise privacy issues. 

\textbf{Cache Structure.}
Using tickets requires to not only maintain an object cache at the server (for static client information -- Fig.\ \ref{fig:signaling}-A), but also a cache of tickets at the side (Fig.\ \ref{fig:signaling}-B). 
For each ticket, this cache retains the type of the cached object, the associated server's identity (i.e., SNI) for scoping (to prevent tracking), and a reference to the matching object, as we support caching multiple objects (e.g., certificate chains). 
On the server side, an object cache is required to retain static client information and link it to tickets. 
As no scoping is required here, a single lookup mechanism via the ticket suffices. 

\textbf{Extension Format.}
For the exchange of information between client and server, we create a dedicated TLS extension, as this allows for minimal modification and ensures interoperability.
More specifically, we define extensions for the respective \textit{Hello} messages of client and server (cf.\ Sec.\ \ref{sec:background}).
On both sides, we utilize a single list, such that the extension consists of the four-byte header, a two byte length field, and the (potentially empty) list. 
The \textit{Client Hello} message then contains a list of entries with the form object type (1 B), ticket length (1 B) and ticket (see below), where object types are chosen according to RFC 7924 \cite{Santesson2016Transport}. 
Similarly, the \textit{Server Hello} message contains a list of elements with the form object type (1 B) and ticket index (2 B), indicating the server's selection of cached objects. 
In case the server does not cache any of the objects signaled by the client, it responds with an empty list.
During the initial handshake, client and server signal support for \name{} using a zero-length extension.

\textbf{Ticket Generation.}
As we use tickets to identify cache objects, client and server need to obtain the same ticket when caching static client information.
Typically, e.g., in session resumption \cite{Eronen2008Transport}, one party generates a ticket and explicitly transmits this to the other party.
However, this results in unnecessary bandwidth overhead.
Instead, \name{} relies on implicit ticket generation based on the TLS key derivation function \cite{rfcTLS12} to derive eight-byte tickets from the \textit{MasterSecret}: 
\texttt{PRF(MasterSecret, "ssc\_ticket\_label", object type || object hash)}. 
As ownership of the actual cached object is still proven in the handshake, no substantial bit security needs to be provided by the ticket's length. 
Furthermore, our mechanism generates fresh tickets for each connection, thus cryptographically decoupling connections. 

\textbf{Adapted Behavior.}
To realize our caching scheme for static client information, we adjust different steps of the TLS handshake at the client and server.
First, when building the \textit{Client Hello} message, the client checks its cache for available tickets scoped to the server and includes them into the extension (if no tickets are available, it sends an empty extension to signal support for \name{}).
Upon receiving tickets from the client, the server checks its cache for corresponding entries and responds with the cache entry it selected (or an empty list if no matching cache entry exists).
If the client only sends an empty extension (to signal support for \name{}), the server likewise signals its support.
Should either client or server not support \name{}, the handshake proceeds as normal, thus ensuring backwards compatibility.
Upon receiving the \textit{Server Hello} message, the client keeps track of server support for \name{} and parses the server's selection of cached objects.
If a cached certificate should be used, the client replaces the content of the \textit{Certificate} message with a zero-length message. 
The server expects such an abbreviated message and uses the selected cached certificate chain instead. 
Finally, after exchanging the \textit{Finished} messages, both peers generate (new) tickets (as described above) and update their caches. 

By caching static client information, we can omit redundant content of messages sent by the client, most notably the client certificate, and thus substantially reduce bandwidth usage.

\subsection{Minimizing Processing with Cached Static Information} \label{sec:des:cache}

Besides cutting bandwidth consumption, \name{} also leverages cached information to substantially reduce the processing overhead for validating certificate chains.
To this end, \name{} validates the certificate chain only once during the initial handshake and subsequently only re-validates time-dependent operations, e.g., certificate revocation checks.
Thus, \name{} not only substantially reduces processing overhead but also obviates the need to cache the complete certificate chain and thus allows to save valuable memory on constrained devices.
In an essence, only the public key of the leaf certificate required during every handshake as well as information required to perform time dependent validation steps need to be cached. 

These steps validate certificate lifetimes, for which it suffices to cache the two most restrictive timestamps (i.e., the maximum of ``not before'' and minimum of ``not after'' values across the certificate chain).
Furthermore, if optional checks for revocation using Certificate Revocation Lists (CRL) \cite{Boeyen2008Internet} or Online Certificate Status Protocol (OCSP) \cite{Santesson2013X.509} are used, also the hash of the issuer name and public key, the serial number, and potentially the CRL distribution points are cached. 

Thus, by substantially reducing the number of signature verifications for repeated handshakes, \name{} speeds up handshakes and cuts energy consumption.
Likewise, as cache entries are reduced to one public key and two timestamps (if no optional revocation checks are used) instead of storing complete certificate chains, the memory overhead of \name{} is well-manageable even for memory constrained IIoT devices. 

\subsection{Interoperability with TLS 1.3} \label{sec:des:tls13}

While TLS 1.2 still is the prevalent choice in IIoT scenarios~\cite{dahlmanns2022tls}, future deployments will likely only support TLS 1.3 and should equally benefit from the tremendous improvements of \name{}.
As TLS 1.3 introduces significant changes to connection establishment (cf.\ Sec.\ \ref{sec:background}), we briefly discuss how these changes impact \name{}. %
The most notable change in TLS 1.3 is that all messages after the \textit{Hello} messages are encrypted and an additional \textit{Encrypted Extension} message is introduced. 
We leverage this to move the server-side signaling to this message for additional tracking protection. 

The changes to the \textit{Certificate} message require further consideration.
Here, TLS 1.3 adds a request context and the option to add extensions to each certificate in the chain. 
Although providing useful functionality, this might lead to non-static content, complicating caching. 
To address this, we leverage a conversion function that (after verification of the certificate chain) transforms a TLS 1.3 message into its TLS 1.2 equivalent by removing certificate extensions and request context, which can then be handled by \name{} as for TLS 1.2.
An abbreviated \textit{Certificate} message is then achieved by utilizing zero-length certificate fields followed by their non-static extensions rather than a zero-length \textit{Certificate} message. 

With these minor adaptations, the advantages of \name{} also apply to constrained IIoT devices using TLS 1.3.

\section{Evaluation} \label{sec:eval}

To quantify the improvements of \name{} and thus the potential to widely deploy TLS in constrained IIoT environments, we implement \name{} on top of wolfSSL version 5.6.0 \cite{wolfSSL}. 
We implement RFC 7924~\cite{Santesson2016Transport}, our caching scheme for static client information, and our approach to reduce processing overhead based on caching for TLS 1.2 and 1.3.

\subsection{Bandwidth Improvements} \label{sec:eval:bw}

To demonstrate the effectiveness of \name{} in reducing the bandwidth overhead of TLS handshakes by omitting redundant transmissions of certificate chains, we perform measurements of bandwidth usage across various parameters.

\textbf{Methodology.}
We evaluate \name{} in a Docker \cite{Docker} environment by deploying both a client and server utilizing our adapted wolfSSL library in a docker container based on Alpine Linux version 3.18.3.
A third docker container runs \texttt{tcpdump} \cite{tcpdump} to capture communication over a virtual Ethernet network.
As wireless networks used in constrained IIoT environments impose tight restrictions on message sizes, TLS messages often get fragmented \cite{rademacher2022bounds, bodenhausen2023securing}.
Thus, we adapt the virtual network to study different Maximum Transmission Units (MTUs):
127 B (MTU in IEEE 802.15.4 wireless networks~\cite{rfc9006}), 576 B (minimum receivable datagram size IPv4 devices must support~\cite{1981Internet}), and 1500 B (typical MTU in Ethernet~\cite{2022IEEE}).

To put the bandwidth improvements of \name{} into perspective, we compare
\begin{enumerate*}[label=(\roman*)]
\item an unmodified version of wolfSSL without caching support (\emph{Vanilla}),
\item our extended version of wolfSSL with support for RFC 7924 (\emph{RFC7924}),
\item our implementation of \name{} (\emph{\name{}}), and
\item an unmodified version of wolfSSL with ID-based (TLS 1.2) or ticket-based (TLS 1.3) session resumption (\emph{Sess.Res.}).
\end{enumerate*}

For each measurement, we establish a connection, send one message per direction and then close the connection. 
Unless stated otherwise, both peers use a chain of three 2048-bit RSA certificates for authentication.
Root certificates are not transmitted, as allowed by the specification \cite{rfcTLS12}.
Connections relying on caching or resumption are preceded by an initial full handshake to populate the cache resp.\ generate a session.
We repeat each measurement ten times, report the arithmetic mean over these measurements, and indicate the minimum and maximum observed values through error bars.

\begin{figure}[t]
    \centering
    \includegraphics[width=1\columnwidth]{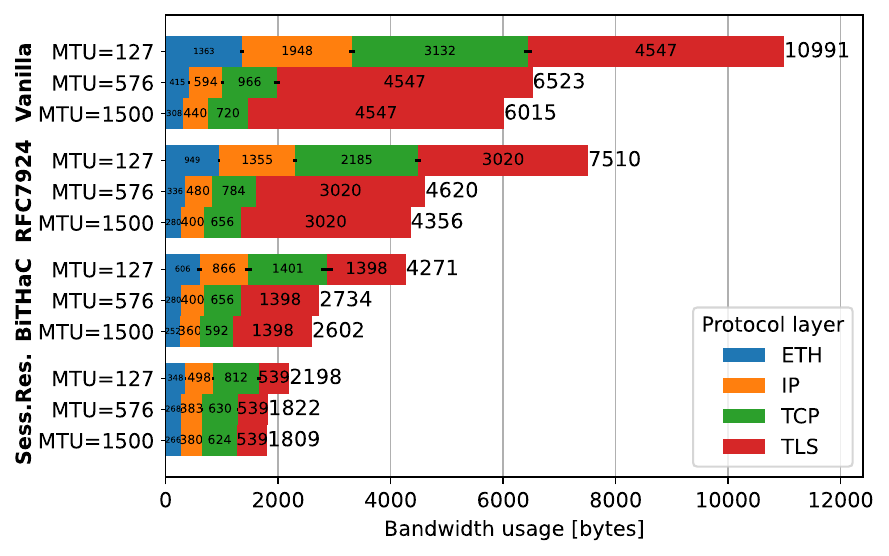}
    \vspace{-2em}
    \caption{
        \name{} substantially reduces bandwidth usage compared to unmodified TLS 1.2 as well as straightforward caching based on RFC 7924.
        While session resumption provides even more savings, those weaken security. 
        }
    \label{fig:eval-bw-layers}
\end{figure}

\textbf{Different MTUs.}
In Fig.\ \ref{fig:eval-bw-layers}, we report on the resulting bandwidth usage of the four TLS configurations for TLS 1.2 and varying MTUs, divided by the different protocol layers.
As expected, the unmodified TLS configuration (Vanilla) consumes the most bandwidth with 6 to 11 kB (slight deviations denoted by the error bars result from race conditions leading to different numbers of acknowledgments).
Already by deploying RFC 7924 to cache static server information, bandwidth usage can be reduced by 31.67\% to 27.58\% (depending on the MTU).
When using \name{} to additionally cache static client information, this can be further decreased to between 43.25\% and 38.86\%.
While session resumption is even more efficient and decreases the required bandwidth by up to 80\%, this comes at the cost of cryptographically linking \emph{all} resumed sessions to the initial session, severely limiting the allowed lifetime of cached sessions ($<$ 24 h for TLS 1.2 sessions~\cite{rfcTLS12}), thus requiring to carry out full handshakes frequently~\cite{rademacher2022bounds}.
Still, \name{} and session resumption are not mutually exclusive and can be used in tandem to combine their strengths. 

Considering the influence of varying MTUs, we observe that the different network layers contribute differently.
While the bandwidth required for TLS remains unchanged, a decrease in the MTU leads to a disproportionate increase in bandwidth overhead on the lower layers.
This particularly highlights the importance of optimizing bandwidth consumption for severely constrained wireless networks such as IEEE 802.15.4.

\textbf{Savings of \name{}.}
Further investigating the bandwidth reductions realized by \name{}, we observe that the server's \textit{Certificate} message is reduced to a constant size of 42 B with caching.
Likewise, the variable size of the client's \textit{Certificate} message is reduced to a constant size of only 9 B, due to \name{}'s more efficient caching scheme for static client information. 
Furthermore, the \textit{CertificateRequest} message is replaced with the selected fingerprint, which results in a marginal reduction of the already constant size of 45 B to 42 B.
To realize these savings, \name{} introduces a reasonable overhead of combined 111 B in the \textit{Hello} messages of client and server. 
For initial handshakes, \name{}'s signaling mechanism (cf.\ Sec.\ \ref{sec:des:server}) causes merely 8 B of overhead.

Notably, \name{} replaces variable-sized static information with a fixed, small placeholder.
Consequently, for larger certificates (or chains), bandwidth savings become even more pronounced.
E.g., when switching to a 4096-bit RSA certificate chain for the server, \name{} reduces the TLS portion of the bandwidth usage from 5667 B to only 1494 B.
Thus, with the ongoing switch to post-quantum cryptography, \name{}'s optimizations become even more relevant.

\begin{figure}[t]
    \centering
    \includegraphics[width=1\columnwidth]{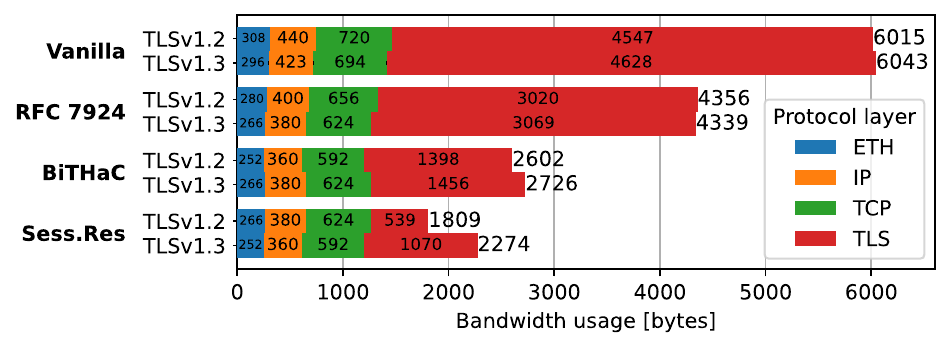}
    \vspace{-2em}
    \caption{
        When switching from TLS 1.2 to 1.3, \name{} is still able to achieve substantial bandwidth savings.
        In contrast, session resumption cannot uphold the same level of bandwidth savings due to the switch to session tickets.
    }
    \label{fig:eval-bw-layers-tls13}
\end{figure}

\textbf{TLS 1.2 vs.\ 1.3.}
In Fig.\ \ref{fig:eval-bw-layers-tls13}, we compare TLS 1.2 and 1.3 for an MTU of 1500 B.
Overall, the results are mostly comparable, with \name{} realizing a bandwidth reduction of more than 3.3 kB (54.89\%) for TLS 1.3.
Particularly noteworthy is the relative improvement of \name{} compared to session resumption, which only realizes additional bandwidth savings of 16.58\% (452 B) for TLS 1.3, with still weaker security guarantees.
Session resumption's increase in bandwidth mainly stems from the switch to session tickets in TLS 1.3 to provide PFS \cite{rfcTLS13}.
Consequently, with the ongoing shift towards TLS 1.3 in IIoT scenarios~\cite{dahlmanns2022tls}, the added security of \name{} over session resumption becomes even more attractive.

\subsection{Processing Improvements} \label{sec:eval:speed}

Besides substantially reducing bandwidth overhead, \name{} also reduces processing overhead using cached static information to avoid redundant computations (cf.\ Sec.\ \ref{sec:des:cache}).
Most notably, for cached certificate chains, this promises to save several costly signature verifications.

\textbf{Methodology.}
To evaluate the processing advantages of \name{}, we implement the corresponding optimizations on the client side for wolfSSL \cite{wolfSSL} version 5.7.0 and perform a series of measurements on a Raspberry Pi Zero W equipped with a 1GHz single-core CPU. 
In our measurement setup, client and server communicate directly via the local interface.
We record the total E2E duration of handshakes, repeat measurements 100 times, and report the arithmetic mean as well as the standard deviation over these repetitions.

\begin{figure}[t]
    \centering
    \includegraphics[width=1\columnwidth]{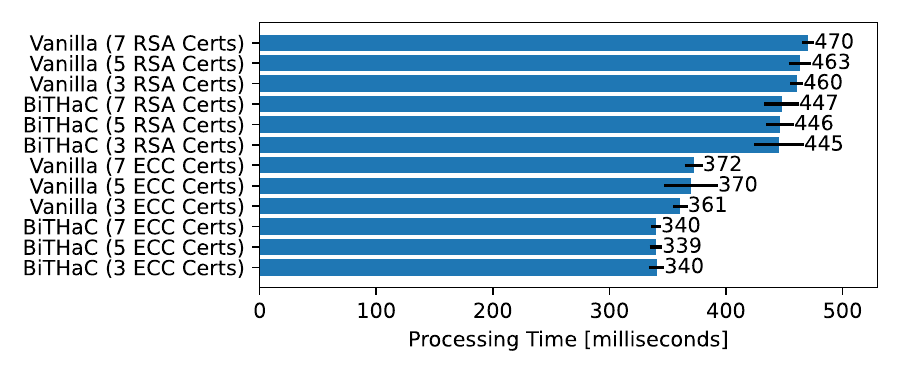}
    \vspace{-2em}
    \caption{
        \name{} reduces the variable processing overhead for validating the entire certificate chain to a constant overhead that only depends on the public key in the leaf certificate, achieving a reduction of up to 8.6\% (32 ms). 
    }
    \label{fig:processing}
\end{figure}

\textbf{Savings of \name{}.}
In Fig.\ \ref{fig:processing}, we report on the processing improvements of \name{} compared to unmodified TLS. 
As expected, connection establishments with RSA certificates invoke a larger processing overhead than ECC certificates and longer certificate chains cause additional overhead. 
For \name{}, however, we measure a constant processing overhead for both certificate types. 
For RSA, this leads to a reduction of 3.3\% (15 ms) to 4.9\% (23 ms), while \name{} achieves an even larger relative reduction for the already more efficient case of ECC certificate chains, were the overhead is reduced by 5.5\% (21 ms) to 8.6\% (32 ms). 
Thus, \name{} not only realizes substantial bandwidth improvements, but additionally speeds-up handshakes by tens of ms, which constitutes a significant improvement for latency-critical IIoT applications~\cite{hiller2018secure,michaelides2025latency}.

\subsection{Memory Costs} \label{sec:eval:memory}

While \name{} realizes substantial bandwidth and processing savings, these require to cache additional information.
To assess whether this is feasible for constrained devices, we evaluate the memory overhead of \name{}.

\textbf{Methodology.}
The evaluation setup is similar to the bandwidth evaluation (cf.\ Sec.\ \ref{sec:eval:bw}).
However, instead of capturing network traffic, we utilize the memory profiler Bytehound \cite{Bytehound}.
Furthermore, we use the \textit{lowresource} compilation option of wolfSSL to obtain a realistic baseline for constrained devices. 
With this setup, we record the maximum measured memory overhead for various events of the connection setup.

\textbf{Peak Memory Usage.}
First, we observe that the memory overhead usually peaks while sending the \textit{ClientKeyExchange} message. 
Moreover, the overhead for a single cached certificate, i.e., a client connecting only to one particular server, is almost negligible small.
Specifically, the relative memory overhead amounts to 3.44\% on the initial handshake and to 1.22\% for subsequent handshakes. 
Thus, the memory overhead of \name{} when connecting to only one server is negligible and well-manageable even for tightly-constrained devices.

\begin{figure}[t]
    \centering
    \includegraphics[width=1\columnwidth]{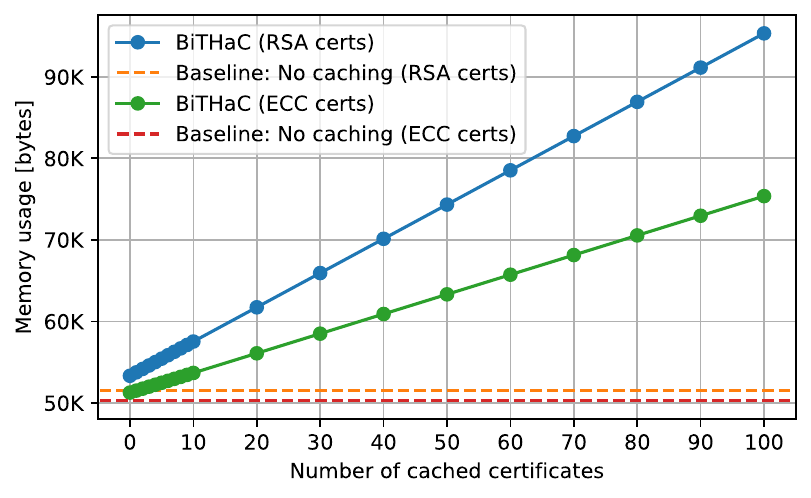}
    \vspace{-2em}
    \caption{
        The memory usage of \name{} scales linearly in the amount of cached certificates and is well-manageable for memory-constrained IIoT devices.
    }
    \label{fig:eval-bw-mem-2}
\end{figure}

\textbf{Memory Usage for Larger Caches.}
To assess the memory usage based on the number and type of cached certificates, we examine the memory overhead for different types (2048-bit RSA keys and 256-bit ECC keys) and numbers of cached certificates, again for chains of length 3. 
As shown in Fig.~\ref{fig:eval-bw-mem-2}, memory usage scales linearly.
More precisely, each cache entry amounts to an overhead of 420 B for RSA certificates and 241 B for ECC certificates.
Considering that constrained IIoT devices connect to a small number of servers, the substantial bandwidth and processing savings clearly outweigh the resulting minor memory overhead.
Furthermore, if cache size should become a limiting factor, old entries can be pruned.

Overall, \name{} reduces both bandwidth consumption and computational overhead of TLS handshakes, while incurring only minor costs in terms of a memory overhead.

\section{Discussion} \label{sec:discussion}

\name{} substantially reduces bandwidth \emph{and} computational overhead without impacting E2E security.
By refraining from protocol-breaking changes, we ensure interoperability with legacy deployments. 
As both peers are authenticated and no trust in a third party is needed, strong security is assured.  

\textbf{Does \name{} weaken security?}
In contrast to other approaches such as offloading~\cite{Fouladgar2006Tiny,Polk2007Server,Hummen2014Delegation,Raza2016S3K}, which require a trusted third party, or improvements at the cost of security, e.g., by foregoing perfect forward secrecy in ID-based session resumption \cite{rfcTLS12,Gupta2005Sizzle,Sobh2008Performance}, \name{} shrinks TLS handshakes while preserving security guarantees. 
Most importantly, \name{} does not change the semantics of the key establishment of the handshake.
Merely the explicit transmission of bandwidth-heavy parts, particularly the certificate chain, is omitted. 
Still, those remain cryptographically linked to the handshake through the included tickets or fingerprints.
Furthermore, both peers still prove possession of the associated private keys.
Our optimized certificate cache warrants additional consideration, as the certificate chain is stored only in significantly abbreviated form (cf.\ Sec.\ \ref{sec:des:cache}). 
To rule out potential attack vectors, only public keys included in a verified certificate chain are added to the cache after an initial successful handshake.
Thus, attack potential only arises from an attacker using a certificate chain that was compromised after it was added to the cache or through direct cache modification. 
As \name{} upholds support for revocation checks, the risks arising from compromised certificate chains are identical to TLS without caching.
Likewise, direct manipulation of the cache requires an attacker to modify application memory, which also contains further security-critical information such as the session key, resulting in comparable risk to plain TLS.

\textbf{Is there an increased risk for device tracking?}
Since \name{} caches certificates, which hold identifiable information, the question of privacy risks arises. 
However, these risks are not larger than those inherent to client certificates~\cite{foppe2018exploiting}. 
For our discussion, we differentiate between TLS 1.2 \cite{rfcTLS12}, where certificates are transmitted in clear, thus allowing for passive tracking, and TLS 1.3 \cite{rfcTLS13}, where additional encryption and changes in the handshake prevent passive tracking (cf.\ Sec.\ \ref{sec:back:tls}). 
Due to these stronger privacy guarantees, we focus our discussion on TLS 1.2.
To prevent tracking, \name{} scopes the cached client-side information (cf.\ Sec.\ \ref{sec:design}), i.e., the client onlys transmit fingerprints and tickets associated with the current peer. 
Without proper scoping, e.g., na\"ively sending all fingerprints in the cache to every server, clients would become sufficiently identifiable for tracking. 
For server-side caching, tickets are freshly generated after each connection and implicitly derived.
Hence, passive correlation of TLS connections is prevented even though tickets are transmitted in clear. %
As this does impact the additional privacy guarantees of TLS 1.3., \name{} does not add any new risks over those deemed an acceptable trade-off for the use of TLS anyways.

\textbf{What about compatibility with legacy devices?}
In contrast to other approaches to reduce the overhead of TLS which rely on protocol breaking changes (cf.\ Sec.\ \ref{sec:rel_work}), \name{} ensures compatibility with legacy devices by using the built-in extension mechanism of TLS \cite{rfcTLS12, rfcTLS13}.
Both peers signal support for \name{} by including a corresponding extension in their \textit{Hello} message (cf.\ Sec.\ \ref{sec:des:server}).
Including arbitrary (new) extensions is fully protocol compliant and a peer will simply ignore unsupported extensions~\cite{rfcTLS12, rfcTLS13}.
Thus, if one peer does not signal support for \name{}, the connection establishment continues with a default TLS handshake.

\textbf{How is \name{} different from RFC 7924?}
\name{} builds upon the cached information extension (RFC 7924) \cite{Santesson2016Transport}.
However, RFC 7924 merely targets to reduce the bandwidth overhead resulting from the use of \emph{server} certificates and \textit{CertificateRequest} messages and leaves cache structure and content to individual implementations.
In contrast, \name{} not only adds functionality to significantly reduce the bandwidth overhead resulting from \emph{client} certificates (whose use is often required in IIoT scenarios~\cite{rademacher2022bounds,heimgaertner2018distributed}), but additionally substantially reduces the computational overhead resulting from the verification of TLS certificates on constrained IIoT devices.
As such, \name{} even provides performance improvements over RFC 7924 when only using server authentication.

\textbf{Why not use session resumption instead?}
The advantages of the full authentication provided by \name{} over ID-based session resumption used up to TLS 1.2 \cite{rfcTLS12}, where keys for resumed sessions are directly derived from previous secrets, are striking due to the resulting lack of PFS and limited session lifetime.
For TLS 1.3, which provides PFS,  \name{}'s advantages require further consideration.
Most notably, \name{} performs a full handshake for each connection and thus provides identical security as TLS without session resumption.
Hence, while session resumption has a bandwidth advantage over \name{} (cf.\ Sec.\ \ref{sec:eval:bw}), these savings are limited to rather short session ticket lifetimes.
While TLS 1.3 generally allows for a ticket lifetime of up to 7 days \cite{rfcTLS13}, these are significantly restricted by policies or regulations~\cite{rademacher2022bounds,heimgaertner2018distributed}, down to the upper limit of 24 hours for TLS 1.2~\cite{rfcTLS12}.
As a cache entry in \name{} can remain valid for the complete validity period of cached certificates, its lifetime greatly exceeds that of sessions tickets.
Notably, session resumption and \name{} are not mutually exclusive and can thus be used together to combine short-term (session resumption) \emph{and} long-term (\name{}) savings.

\section{Conclusion} \label{sec:conclusion}

By bi-directionally caching static parts of recurring TLS handshakes, \name{} addresses the substantial bandwidth and processing overhead of TLS, which often prevents its use in constrained IIoT environments.
When leveraging \name{}, redundant information, e.g., certificates, neither have to be transmitted nor corresponding costly computations, such as the validation of certificates, performed repeatedly.
Notably, \name{} is fully compatible with legacy implementations through the use of TLS extensions, allowing for incremental deployability.
Finally, through our approach of extending caching to redundant computations, we particularly improve upon the computational overhead of asymmetric cryptography, while keeping the memory overhead, i.e., the inherent trade-off of any caching mechanism, minimal. 
Our evaluation of \name{} shows bandwidth savings of up to 61.1\% and processing reductions of up to 8.5\%, while the memory overhead remains well-manageable.
Most notably and unlike other approaches such as session resumption, \name{} fully upholds the strong security notions of E2E security.
With the ongoing deployment of post-quantum cryptography in TLS and resulting further increases in bandwidth demand, the savings offered by \name{} will become even more relevant. 

\section*{Acknowledgments}

Funded by the Deutsche Forschungsgemeinschaft (DFG, German Research Foundation) under Germany's Excellence Strategy -- EXC-2023 Internet of Production -- 390621612, the German Federal Office for Information Security (BSI) under project funding reference number 01MO23003D (PlusMoSmart), and the German Federal Ministry of the Interior represented by the German Federal Agency for Public Safety Digital Radio (BDBOS) under project funding reference number 16BEC0049 (MissionXconnect).
The responsibility for the content of this publication lies with the authors.


\begin{thebibliography}{10}
\providecommand{\url}[1]{#1}
\csname url@samestyle\endcsname
\providecommand{\newblock}{\relax}
\providecommand{\bibinfo}[2]{#2}
\providecommand{\BIBentrySTDinterwordspacing}{\spaceskip=0pt\relax}
\providecommand{\BIBentryALTinterwordstretchfactor}{4}
\providecommand{\BIBentryALTinterwordspacing}{\spaceskip=\fontdimen2\font plus
\BIBentryALTinterwordstretchfactor\fontdimen3\font minus
  \fontdimen4\font\relax}
\providecommand{\BIBforeignlanguage}[2]{{%
\expandafter\ifx\csname l@#1\endcsname\relax
\typeout{** WARNING: IEEEtran.bst: No hyphenation pattern has been}%
\typeout{** loaded for the language `#1'. Using the pattern for}%
\typeout{** the default language instead.}%
\else
\language=\csname l@#1\endcsname
\fi
#2}}
\providecommand{\BIBdecl}{\relax}
\BIBdecl

\bibitem{kim2017smart}
T.-h. Kim, C.~Ramos, and S.~Mohammed, ``{Smart City and IoT},'' \emph{Future
  Generation Computer Systems}, vol.~76, 2017.

\bibitem{zhang2020vehicle}
H.~Zhang and X.~Lu, ``{Vehicle communication network in intelligent
  transportation system based on Internet of Things},'' \emph{Computer
  Communications}, vol. 160, 2020.

\bibitem{ko2011industry}
J.~Ko, J.~Eriksson, N.~Tsiftes, S.~Dawson-Haggerty, J.-P. Vasseur, M.~Durvy,
  A.~Terzis, A.~Dunkels, and D.~Culler, ``{Industry: Beyond interoperability:
  Pushing the performance of sensor network IP stacks},'' in \emph{SenSys},
  2011.

\bibitem{lenz2025cofacs}
S.~Lenz, D.~Schachtschneider, S.~Jonas, L.~Tirpitz, S.~Geisler, and M.~Henze,
  ``{CoFacS -- Simulating a Complete Factory to Study the Security of
  Interconnected Production},'' in \emph{LCN}, 2025.

\bibitem{rondon2020poisonivy}
L.~P. Rondon, L.~Babun, A.~Aris, K.~Akkaya, and A.~S. Uluagac, ``{PoisonIvy:
  (In) secure Practices of Enterprise IoT Systems in Smart Buildings},'' in
  \emph{BuildSys}, 2020.

\bibitem{wagner2024madtls}
E.~Wagner, D.~Heye, M.~Serror, I.~Kunze, K.~Wehrle, and M.~Henze, ``{Madtls:
  Fine-grained Middlebox-aware End-to-end Security for Industrial
  Communication},'' in \emph{ACM ASIA CCS}, 2024.

\bibitem{dahlmanns2022tls}
M.~Dahlmanns, J.~Lohm{\"o}ller, J.~Pennekamp, J.~Bodenhausen, K.~Wehrle, and
  M.~Henze, ``{Missed Opportunities: Measuring the Untapped TLS Support in the
  Industrial Internet of Things},'' in \emph{ACM ASIA CCS}, 2022.

\bibitem{rademacher2022bounds}
M.~Rademacher, H.~Linka, J.~Konrad, T.~Horstmann, and K.~Jonas, ``{Bounds for
  the Scalability of TLS over LoRaWAN},'' in \emph{ITG MKT}, 2022.

\bibitem{heimgaertner2018distributed}
F.~Heimgaertner and M.~Menth, ``Distributed controller communication in virtual
  power plants using smart meter gateways,'' in \emph{ICE IEEE/ITMC}, 2018.

\bibitem{rfcConstrained}
C.~Bormann, M.~Ersue, and A.~Keranen, ``{Terminology for Constrained-Node
  Networks},'' RFC Editor, RFC 7228, 2014.

\bibitem{henze2016cppl}
M.~Henze, J.~Hiller, S.~Schmerling, J.~H. Ziegeldorf, and K.~Wehrle, ``{CPPL:
  Compact Privacy Policy Language},'' in \emph{WPES}, 2016.

\bibitem{bodenhausen2023securing}
J.~Bodenhausen, C.~Sorgatz, T.~Vogt, K.~Grafflage, S.~R{\"o}tzel,
  M.~Rademacher, and M.~Henze, ``{Securing Wireless Communication in Critical
  Infrastructure: Challenges and Opportunities},'' \emph{MobiQuitous}, 2023.

\bibitem{rademacher2021path}
M.~Rademacher, H.~Linka, T.~Horstmann, and M.~Henze, ``{Path Loss in Urban LoRa
  Networks: A Large-Scale Measurement Study},'' in \emph{VTC Fall}, 2021.

\bibitem{neto2016aot}
A.~L.~M. Neto, A.~L. Souza, I.~Cunha, M.~Nogueira, I.~O. Nunes, L.~Cotta,
  N.~Gentille, A.~A. Loureiro, D.~F. Aranha, H.~K. Patil \emph{et~al.}, ``{AoT:
  Authentication and Access Control for the Entire IoT Device Life-Cycle},'' in
  \emph{SenSys}, 2016.

\bibitem{Schwabe2021More}
P.~Schwabe, D.~Stebila, and T.~Wiggers, ``{{More Efficient Post-quantum KEMTLS
  with Pre-distributed Public Keys}},'' in \emph{ESORICS}, 2021.

\bibitem{bang2022iot}
A.~O. Bang, U.~P. Rao, A.~Visconti, A.~Brighente, and M.~Conti, ``{An IoT
  Inventory Before Deployment: A Survey on IoT Protocols, Communication
  Technologies, Vulnerabilities, Attacks, and Future Research Directions},''
  \emph{Computers \& Security}, vol. 123, 2022.

\bibitem{Gupta2005Sizzle}
V.~Gupta, M.~Wurm, Y.~Zhu, M.~Millard, S.~Fung, N.~Gura, H.~Eberle, and S.~C.
  Shantz, ``{Sizzle: A standards-based end-to-end security architecture for the
  embedded Internet},'' \emph{Pervasive and Mobile Computing}, 2005.

\bibitem{Tschofenig2016Transport}
H.~Tschofenig and T.~Fossati, ``{Transport Layer Security (TLS) / Datagram
  Transport Layer Security (DTLS) Profiles for the Internet of Things},'' RFC
  7925, 2016.

\bibitem{Raza20126LoWPAN}
S.~Raza, D.~Trabalza, and T.~Voigt, ``{6LoWPAN Compressed DTLS for CoAP},'' in
  \emph{DCOSS}, 2012.

\bibitem{Rescorla2023Compact}
E.~Rescorla, R.~Barnes, H.~Tschofenig, and B.~M. Schwartz, ``{Compact TLS
  1.3},'' Internet-Draft, 2023, work in Progress.

\bibitem{Kothmayr2012DTLS}
T.~Kothmayr, C.~Schmitt, W.~Hu, M.~Brünig, and G.~Carle, ``{A DTLS based
  end-to-end security architecture for the Internet of Things with two-way
  authentication},'' in \emph{LCNW - Workshops}, 2012.

\bibitem{Wouters2014Using}
P.~Wouters, H.~Tschofenig, J.~Gilmore, S.~Weiler, and T.~Kivinen, ``{Using Raw
  Public Keys in Transport Layer Security (TLS) and Datagram Transport Layer
  Security (DTLS)},'' RFC 7250, 2014.

\bibitem{Hummen2014Delegation}
R.~Hummen, H.~Shafagh, S.~Raza, T.~Voig, and K.~Wehrle, ``{{Delegation-based
  authentication and authorization for the IP-based Internet of Things}},'' in
  \emph{SECON}, 2014.

\bibitem{Raza2016S3K}
S.~Raza, L.~Seitz, D.~Sitenkov, and G.~Selander, ``{S3K: Scalable Security With
  Symmetric Keys—DTLS Key Establishment for the Internet of Things},''
  \emph{TASE}, 2016.

\bibitem{Eronen2008Transport}
P.~Eronen, H.~Tschofenig, H.~Zhou, and J.~A. Salowey, ``{Transport Layer
  Security (TLS) Session Resumption without Server-Side State},'' RFC 5077,
  2008.

\bibitem{Tange2020rTLS}
K.~Tange, D.~Howard, T.~Shanahan, S.~Pepe, X.~Fafoutis, and N.~Dragoni,
  ``{{rTLS: Lightweight TLS Session Resumption for Constrained IoT Devices}},''
  in \emph{Information and Communications Security}, 2020.

\bibitem{Apostolopoulos1999Transport}
G.~Apostolopoulos, V.~Peris, and D.~Saha, ``Transport layer security: how much
  does it really cost?'' in \emph{IEEE INFOCOM '99.}, 1999.

\bibitem{Santesson2016Transport}
S.~Santesson and H.~Tschofenig, ``{Transport Layer Security (TLS) Cached
  Information Extension},'' RFC 7924, 2016.

\bibitem{hebrok2023we}
S.~Hebrok, S.~Nachtigall, M.~Maehren, N.~Erinola, R.~Merget, J.~Somorovsky, and
  J.~Schwenk, ``{We Really Need to Talk About Session Tickets: A Large-Scale
  Analysis of Cryptographic Dangers with TLS Session Tickets},'' in
  \emph{USENIX Security 23}, 2023.

\bibitem{rfcTLS12}
T.~Dierks and E.~Rescorla, ``{The Transport Layer Security (TLS) Protocol
  Version 1.2},'' RFC Editor, RFC 5246, 2008.

\bibitem{rfcTLS13}
E.~Rescorla, ``{The Transport Layer Security (TLS) Protocol Version 1.3},'' RFC
  Editor, RFC 8446, 2018.

\bibitem{kampanakis2022faster}
P.~Kampanakis and M.~Kallitsis, ``{Faster post-quantum TLS handshakes without
  intermediate CA certificates},'' in \emph{CSCML}, 2022.

\bibitem{ietf-tls-hybrid-design-10}
D.~Stebila, S.~Fluhrer, and S.~Gueron, ``{Hybrid key exchange in TLS 1.3},''
  IETF, Internet-Draft draft-ietf-tls-hybrid-design-10, Apr. 2024, work in
  Progress.

\bibitem{eggert2014sensorcloud}
M.~Eggert, R.~H{\"a}u{\ss}ling, M.~Henze, L.~Hermerschmidt, R.~Hummen,
  D.~Kerpen, A.~Navarro~P{\'e}rez, B.~Rumpe, D.~Thi{\ss}en, and K.~Wehrle,
  ``{SensorCloud: Towards the Interdisciplinary Development of a Trustworthy
  Platform for Globally Interconnected Sensors and Actuators},'' in
  \emph{Trusted Cloud Computing}, 2014.

\bibitem{serror2021challenges}
M.~Serror, S.~Hack, M.~Henze, M.~Schuba, and K.~Wehrle, ``{Challenges and
  Opportunities in Securing the Industrial Internet of Things},'' \emph{IEEE
  TII}, vol.~17, no.~5, 2021.

\bibitem{karlof2004tinysec}
C.~Karlof, N.~Sastry, and D.~Wagner, ``{TinySec: A Link Layer Security
  Architecture for Wireless Sensor Networks},'' in \emph{SenSys}, 2004.

\bibitem{henze2013maintaining}
M.~Henze, R.~Hummen, R.~Matzutt, D.~Catrein, and K.~Wehrle, ``{Maintaining User
  Control While Storing and Processing Sensor Data in the Cloud},''
  \emph{IJGHPC}, 2013.

\bibitem{michaelides2025industry5G}
S.~Michaelides, S.~Lenz, T.~Vogt, and M.~Henze, ``{Secure Integration of 5G in
  Industrial Networks: State of the Art, Challenges and Opportunities},''
  \emph{Future Generation Computer Systems}, vol. 166, 2025.

\bibitem{brighente2022interference}
A.~Brighente, J.~Mohammadi, P.~Baracca, S.~Mandelli, and S.~Tomasin,
  ``{Interference Prediction for Low-Complexity Link Adaptation in Beyond 5G
  Ultra-Reliable Low-Latency Communications},'' \emph{IEEE Transactions on
  Wireless Communications}, vol.~21, no.~10, 2022.

\bibitem{restuccia2020low}
G.~Restuccia, H.~Tschofenig, and E.~Baccelli, ``{Low-Power IoT Communication
  Security: On the Performance of DTLS and TLS 1.3},'' in \emph{IFIP PEMWN},
  2020.

\bibitem{bodenhausen2024adaptive}
J.~Bodenhausen, L.~Grote, M.~Rademacher, and M.~Henze, ``{Adaptive Optimization
  of TLS Overhead for Wireless Communication in Critical Infrastructure},''
  \emph{CSNet}, 2024.

\bibitem{kothmayr2011poster}
T.~Kothmayr, W.~Hu, C.~Schmitt, M.~Bruenig, and G.~Carle, ``{Poster: Securing
  the Internet of Things with DTLS},'' in \emph{SenSys}, 2011.

\bibitem{brachmann2012end}
M.~Brachmann, S.~L. Keoh, O.~G. Morchon, and S.~S. Kumar, ``{End-to-end
  transport security in the IP-based internet of things},'' in \emph{ICCCN},
  2012.

\bibitem{behrens2017internet}
R.~Behrens and A.~Ahmed, ``{Internet of Things: An end-to-end security
  layer},'' in \emph{ICIN}, 2017.

\bibitem{sahraoui2014compressed}
S.~Sahraoui and A.~Bilami, ``{Compressed and distributed host identity protocol
  for end-to-end security in the IoT},'' in \emph{NGNS}, 2014.

\bibitem{sahraoui2015efficient}
------, ``{Efficient HIP-based approach to ensure lightweight end-to-end
  security in the internet of things},'' \emph{Computer Networks}, vol.~91,
  2015.

\bibitem{granjal2017adaptable}
J.~Granjal and E.~Monteiro, ``{Adaptable End-To-End Security For Mobile IoT
  Sensing Applications},'' in \emph{SafeThings}, 2017.

\bibitem{plusquellic2023privacy}
J.~Plusquellic, E.~E. Tsiropoulou, and C.~Minwalla, ``{Privacy-Preserving
  Authentication Protocols for IoT Devices Using the SiRF PUF},'' \emph{IEEE
  Transactions on Emerging Topics in Computing}, vol.~11, no.~4, 2023.

\bibitem{Raza2013Lithe}
S.~Raza, H.~Shafagh, K.~Hewage, R.~Hummen, and T.~Voigt, ``{Lithe: Lightweight
  Secure CoAP for the Internet of Things},'' \emph{IEEE Sensors Journal}, 2013.

\bibitem{Banerjee2017eeDTLS}
U.~Banerjee, C.~Juvekar, S.~H. Fuller, and A.~P. Chandrakasan, ``{{eeDTLS:
  Energy-Efficient Datagram Transport Layer Security for the Internet of
  Things}},'' in \emph{GLOBECOM}, 2017.

\bibitem{Boeyen2008Internet}
S.~Boeyen, S.~Santesson, T.~Polk, R.~Housley, S.~Farrell, and D.~Cooper,
  ``{Internet X.509 Public Key Infrastructure Certificate and Certificate
  Revocation List (CRL) Profile},'' RFC 5280, 2008.

\bibitem{McGrew2010Compressed}
D.~McGrew and M.~Pritikin, ``{The Compressed X.509 Certificate Format},''
  Internet-Draft, 2010, expired.

\bibitem{Ghedini2020TLS}
A.~Ghedini and V.~Vasiliev, ``{TLS Certificate Compression},'' RFC 8879, 2020.

\bibitem{OrtizYepes2015Optimizing}
D.~A. Ortiz-Yepes, ``{Optimizing TLS for Low Bandwidth Environments},'' in
  \emph{Foundations and Practice of Security}, 2015.

\bibitem{Hoeglund2020PKI4IoT}
J.~Höglund, S.~Lindemer, M.~Furuhed, and S.~Raza, ``{PKI4IoT: Towards public
  key infrastructure for the Internet of Things},'' \emph{Computers \&
  Security}, 2020.

\bibitem{Mattsson2023CBOR}
J.~P. Mattsson, G.~Selander, S.~Raza, J.~Höglund, and M.~Furuhed, ``{CBOR
  Encoded X.509 Certificates (C509 Certificates)},'' Internet-Draft, 2024, work
  in Progress.

\bibitem{Jung2009SSL}
W.~Jung, S.~Hong, M.~Ha, Y.-J. Kim, and D.~Kim, ``{SSL-Based Lightweight
  Security of IP-Based Wireless Sensor Networks},'' in \emph{AINA Workshops},
  2009.

\bibitem{Tschofenig2023TLS/DTLS}
H.~Tschofenig, T.~Fossati, and M.~Richardson, ``{TLS/DTLS 1.3 Profiles for the
  Internet of Things},'' Internet-Draft, 2023, work in Progress.

\bibitem{3rd2011Transport}
D.~E. {Eastlake 3rd}, ``{Transport Layer Security (TLS) Extensions: Extension
  Definitions},'' RFC 6066, 2011.

\bibitem{Fouladgar2006Tiny}
S.~Fouladgar, B.~Mainaud, K.~Masmoudi, and H.~Afifi, ``{Tiny 3-TLS: A Trust
  Delegation Protocol for Wireless Sensor Networks},'' in \emph{ESAS}, 2006.

\bibitem{Polk2007Server}
T.~Polk, D.~Cooper, R.~Housley, A.~N. Malpani, and T.~Freeman, ``{Server-Based
  Certificate Validation Protocol (SCVP)},'' RFC 5055, 2007.

\bibitem{henze2014trustpoint}
M.~Henze, R.~Hummen, R.~Matzutt, and K.~Wehrle, ``{A Trust Point-based Security
  Architecture for Sensor Data in the Cloud},'' in \emph{Trusted Cloud
  Computing}.\hskip 1em plus 0.5em minus 0.4em\relax Springer, 2014.

\bibitem{Hummen2013Towards}
R.~Hummen, J.~H. Ziegeldorf, H.~Shafagh, S.~Raza, and K.~Wehrle, ``Towards
  viable certificate-based authentication for the internet of things,'' ser.
  HotWiSec '13, 2013.

\bibitem{Sobh2008Performance}
T.~S. Sobh, A.~Elgohary, and M.~Zaki, ``Performance improvements on the network
  security protocols,'' \emph{International Journal of Network Security}, 2008.

\bibitem{Hummen2013Extended}
R.~Hummen, J.~Gilger, and H.~Shafagh, ``{Extended DTLS Session Resumption for
  Constrained Network Environments},'' Internet-Draft, 2013, expired.

\bibitem{Shacham2004Client}
H.~Shacham, D.~Boneh, and E.~Rescorla, ``{Client-Side Caching for TLS},''
  \emph{ACM TISSEC}, 2004.

\bibitem{Langley2010Transport}
A.~Langley, ``{Transport Layer Security (TLS) Snap Start},'' Internet-Draft,
  2010, expired.

\bibitem{schukat2014securing}
M.~Schukat, ``{Securing Critical Infrastructure},'' in \emph{DT}, 2014.

\bibitem{Santesson2013X.509}
S.~Santesson, M.~Myers, R.~Ankney, A.~Malpani, S.~Galperin, and D.~C. Adams,
  ``{X.509 Internet Public Key Infrastructure Online Certificate Status
  Protocol - OCSP},'' RFC 6960, 2013.

\bibitem{wolfSSL}
``{{wolfSSL: Embedded TLS Library for Applications, Devices, IoT, and the
  Cloud}},'' \url{https://github.com/wolfSSL/wolfssl}.

\bibitem{Docker}
``{{Docker: Accelerated Container Application Development}},'' Available
  online: \url{https://www.docker.com/}.

\bibitem{tcpdump}
``{{tcpdump: Command-line packet analyzer}},'' Available online:
  \url{https://www.tcpdump.org/}.

\bibitem{rfc9006}
C.~Gomez, J.~Crowcroft, and M.~Scharf, ``{TCP Usage Guidance in the Internet of
  Things (IoT)},'' RFC 9006, Mar. 2021.

\bibitem{1981Internet}
{Information Sciences Institute, University of Southern California},
  ``{Internet Protocol},'' RFC 791, Sep. 1981.

\bibitem{2022IEEE}
``{IEEE Standard for Ethernet},'' \emph{IEEE Std 802.3-2022 (Revision of IEEE
  Std 802.3-2018)}, vol. IEEE 802.3-2022, pp. 1--7025, 2022.

\bibitem{hiller2018secure}
J.~Hiller, M.~Henze, M.~Serror, E.~Wagner, J.~N. Richter, and K.~Wehrle,
  ``{Secure Low Latency Communication for Constrained Industrial IoT
  Scenarios},'' in \emph{IEEE LCN}, 2018.

\bibitem{michaelides2025latency}
S.~Michaelides, J.~Mucke, and M.~Henze, ``{Assessing the Latency of Network
  Layer Security in 5G Networks},'' in \emph{WiSec}, 2025.

\bibitem{Bytehound}
``{{Bytehound - a memory profiler for Linux}},'' Available online:
  \url{https://github.com/koute/bytehound}.

\bibitem{foppe2018exploiting}
L.~Foppe, J.~Martin, T.~Mayberry, E.~C. Rye, and L.~Brown, ``{Exploiting TLS
  Client Authentication for Widespread User Tracking},'' \emph{PETS}, 2018.

\end{thebibliography}
\end{document}